\documentclass[nonacm, sigconf, screen]{acmart}
\AtBeginDocument{%
  \providecommand\BibTeX{{%
    \normalfont B\kern-0.5em{\scshape i\kern-0.25em b}\kern-0.8em\TeX}}}

\usepackage{graphicx}
\usepackage{float}
\usepackage{subcaption}
\usepackage[font=scriptsize,labelfont=bf]{caption}
\usepackage{amsmath}
\usepackage[colorinlistoftodos]{todonotes}
\begin{document}
\title{Multi-Modal Retrieval using Graph Neural Networks}

\author{Aashish Misraa, Ajinkya Kale, Pranav Aggarwal, Ali Aminian}
%%\authornote{Both authors contributed equally to this research.}
%%\email{trovato@corporation.com}
%%\orcid{1234-5678-9012}
%%\author{G.K.M. Tobin}
%%\authornotemark[1]
%%\email{webmaster@marysville-ohio.com}
\affiliation{
\institution{Adobe Inc. \\
\{amisraa, akale, pranagga, aminian\}@adobe.com}
}

%%\renewcommand{\shortauthors}{Misraa, et al.}


\begin{abstract}
Most real world applications of image retrieval such as Adobe Stock, which is a marketplace for stock photography and illustrations, need a way for users to find images which are both visually (i.e.\ aesthetically)  and conceptually (i.e.\ containing the same salient objects) as a query image. Learning visual-semantic representations from images is a well studied problem for  image retrieval. Filtering based on image concepts or attributes is traditionally achieved with index-based filtering (e.g.\ on textual tags) or by re-ranking after an initial visual embedding based retrieval. In this paper, we learn a joint vision and concept embedding in the same high-dimensional space. This joint model gives the user fine-grained control over the semantics of the result set, allowing them to explore the catalog of images more rapidly. We model the visual and concept relationships as a graph structure, which captures the rich information through node neighborhood. This graph structure helps us  learn multi-modal node embeddings using Graph Neural Networks. We also introduce a novel inference time control, based on selective neighborhood connectivity allowing the user control over the retrieval algorithm. We evaluate these multi-modal embeddings quantitatively on the downstream relevance task of image retrieval on MS-COCO dataset and qualitatively on MS-COCO and an Adobe Stock dataset.

\end{abstract}





\maketitle

\section{Introduction}
eCommerce sites now routinely use visual similarity to improve recall and relevance of search and recommendation results ~\cite{jing2015visual, yang2017visual, zhang2018visual, shankar2017deep}. Visual similarity is particularly useful in domains like fashion and home furnishings where product aesthetics are key but are difficult for consumers to describe in textual queries. The need for image similarity features in eCommerce search is even greater for sites whose business is selling images such as  Adobe Stock\footnote{www.stock.adobe.com/}, Getty Images\footnote{www.gettyimages.com/}, and ShutterStock\footnote{www.shutterstock.com/}. As with eCommerce sites selling physical goods, eCommerce sites selling images need to provide ways to map between user's textual, image and joint text-image queries and the textual, structured, and image data associated with the products in the catalog.

Visual search is a well studied problem ~\cite{datta2008image}.  The subjectivity of the search result quality makes it a difficult problem. This problem is exacerbated for eCommerce sites selling images because the results on the page need to match the semantics of the query but also need to represent the diversity of the images available for purchase. Traditional visual search methods optimize for visual cues and as a result often neglect concept similarity and diversity of results, which are key for eCommerce whole page relevance. In addition,  these production systems look at visual similarity as a separate process prior to concept matching. This has two drawbacks:

\begin{figure*}
\centering
  \includegraphics[width=\textwidth]{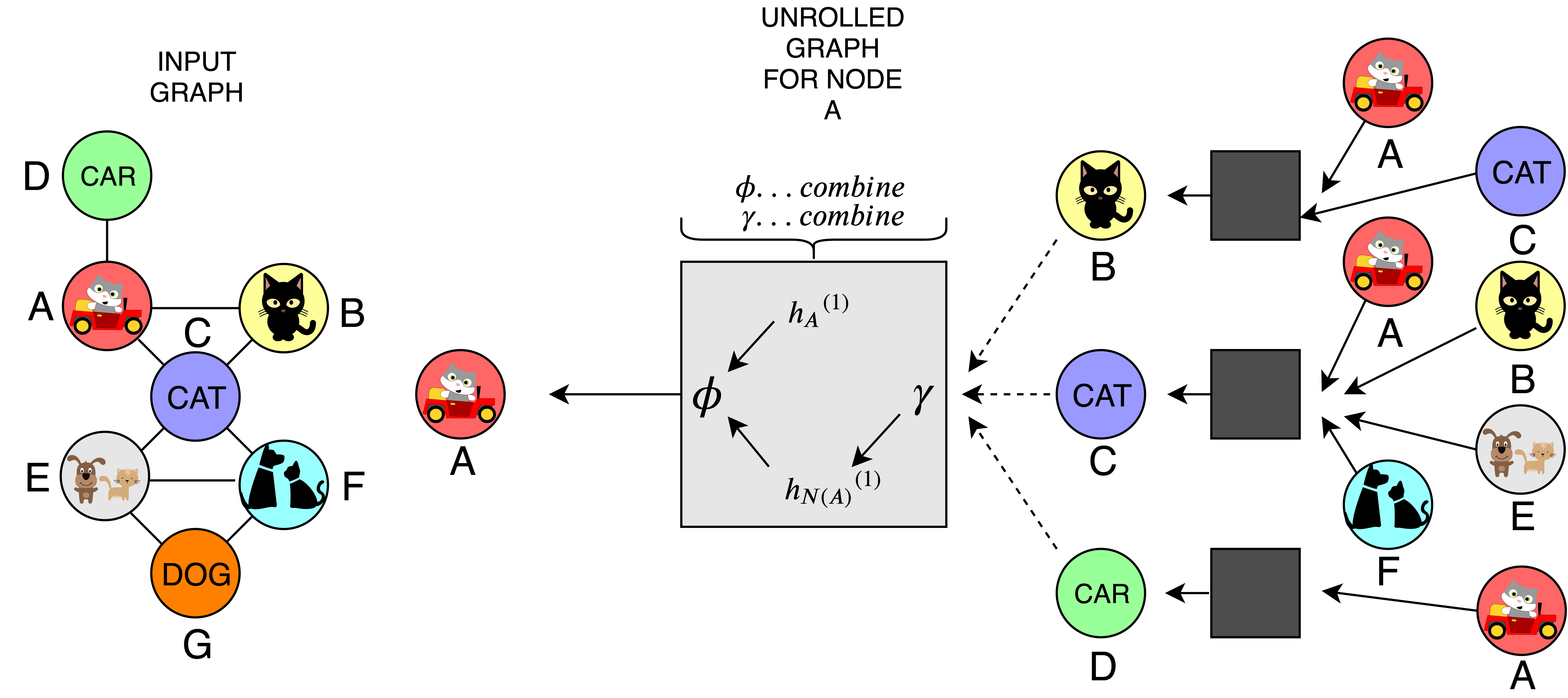}
  \caption{\normalfont Left: Input Graph | Right: Unrolled Graph w.r.t Node A}
  \Description{}
  \label{fig:unrolled-graph}
\end{figure*}

\begin{enumerate}
    \item Concept filtering is bound by the cap on the initial visual similarity retrieval set. As a result, relevant results may be excluded in the initial visual retrieval step. 
    \item Most systems perform ``hard filtering'' of attributes e.g.\ a color filter of {\em tag:red} will remove all images which do not have the relevant tag in their metadata. Given that such metadata is often incomplete, this again results in relevant results being excluded.
\end{enumerate}
These issues are  magnified in a image retrieval system where the user query is text based because user intent is hard to convey with traditional short keyword queries. However, the vast majority of queries on eCommerce sites selling images are keyword based even when image similarity querying is provided.

In this paper we take initial steps towards solving these problems. We explore a joint representation learning framework using Graph Neural Network to model content and visual modalities. This joint representation is achieved via the structure between the two modalities. We create a graph of images and their metadata (in this initial exploration, textual tags) to convert each into a node embedding capturing the structure between the two modalities. We introduce a novel inference system to give users  control over which modality should be weighted more during image retrieval creating  a one-shot retrieval approach instead of the traditional two-step  retrieval with filtering.
Our primary contributions are the following:
\begin{enumerate}
    \item A multi-modal graph embedding framework for visual search
    \item Coarse- and fine-level control over results using inference time dynamic node connectivity for the query node
\end{enumerate}

We provide quantitative and qualitative evaluations to demonstrate the usefulness for our approach for improving the search experience for marketplaces for images like Adobe Stock.  Our approach enables  UX features to aid  search engines in moving beyond binary hard filters to fine-grained soft filters in a continuous space and to allow the user to control  similarity across modalities in a seamless way.

\section{Related Work}
Content-based features for image retrieval use global features to describe the image content by structure\cite{thirtytwo}, shape\cite{thirtythree,thirtyfour,thirtyfive,thirtysix}, color\cite{thirtyseven,thirtyeight} and texture\cite{thirtynine,forty} in a single representation. With the advent of deep learning, it is easier to extract semantic-aware features through the network’s activations from different layers of a classification model. Visual-semantic embeddings are multi-modal joint embeddings which help in tasks like visual-question answering, image captioning and image-caption retrieval. They are primarily obtained in two ways: (1) Canonical Correlation Analysis (CCA) and (2) Metric Learning. CCA-based methods maximize the correlation between linearly projected vectors from two modalities \cite{fiftyeight}. \cite{sixtyone} propose an extension of CCA using a deep learning approach. Metric learning adapts a pairwise real-valued function to the problem at hand by using the training examples as its information source. The method learns an embedding space \cite{sixtythree, sixtyfour} and scales better to large datasets. \cite{sixtythree} combines cross-view ranking constraints with within-view neighborhood structure preservation to generate image-text embeddings. Full sentences, images and image regions were all mapped into a single space by \cite{sixtysix}. \cite{sixtyfour} integrates hard negatives into the ranking loss function to improve retrieval. The loss functions used to execute these approaches generally include triplet loss\cite{sixtyseven} which constraints negative samples to be farther away than positives or variations of it, some of them being Histogram Loss\cite{sixtyeight}, PDDM\cite{sixtynine}, Ladder Loss\cite{four-Ladder} and PATR loss\cite{pranav}. CM-GANs \cite{DBLP:journals/corr/abs-1710-05106} is an another interesting attempt to get images and text modalities in the common learning space by leveraging Generative Adversarial Networks\cite{goodfellow2014generative}. Unlike the above methods, graph based methods are able to capture the multi-modal complexities by using the information present in the node attributes, their graph neighborhood and their connections. This allows better connectivity of information rather than treating each sample independently.

Recently, developments in deep graph structures have made it easier to represent images (nodes) as low-dimensional vector embeddings using their graph neighbourhood information. Graph convolutional neural networks have shown their effectiveness in semi-supervised classification settings ~\cite{DBLP:journals/corr/KipfW16, 10.1145/2806416.2806512, DBLP:journals/corr/GroverL16, 10.1145/2623330.2623732, DBLP:journals/corr/TangQWZYM15, 10.1145/2939672.2939753}. But graph node embeddings have mostly been transductive in nature which makes them hard to generalise to unseen queries.  \cite{DBLP:journals/corr/YangCS16} proposed an architecture that added inductive capabilities, but lacked the graph structure information during inference. GraphSAGE\cite{GraphSAGE} was able to overcome this limitation and generalize new vector embeddings for unseen nodes more efficiently using trainable neural network aggregators. This allow graphs to be more expandable and adaptable to the changing dataset and social trends. Unlike GraphSAGE, GAT~\cite{velikovi2017graph} tries to use all the neighbours of the node to contribute to the target node along with generating self-attention scores on the full 1-hop neighbourhood. We leverage the architecture of GraphSAGE for it's industrial application friendly architecture to build a multi-modal setting which gives more control to customers during search inference. 

\section{Method}
We build on top of GraphSAGE's \cite{GraphSAGE} idea of context construction and information aggregation in a graph neighborhood. We extend it to the multi-modal context and a fine-grained inference control using dynamic node connections controlling the information flow.

\subsection{Graph Structure}
\begin{figure}
    \centering
    \includegraphics[width=0.4\linewidth]{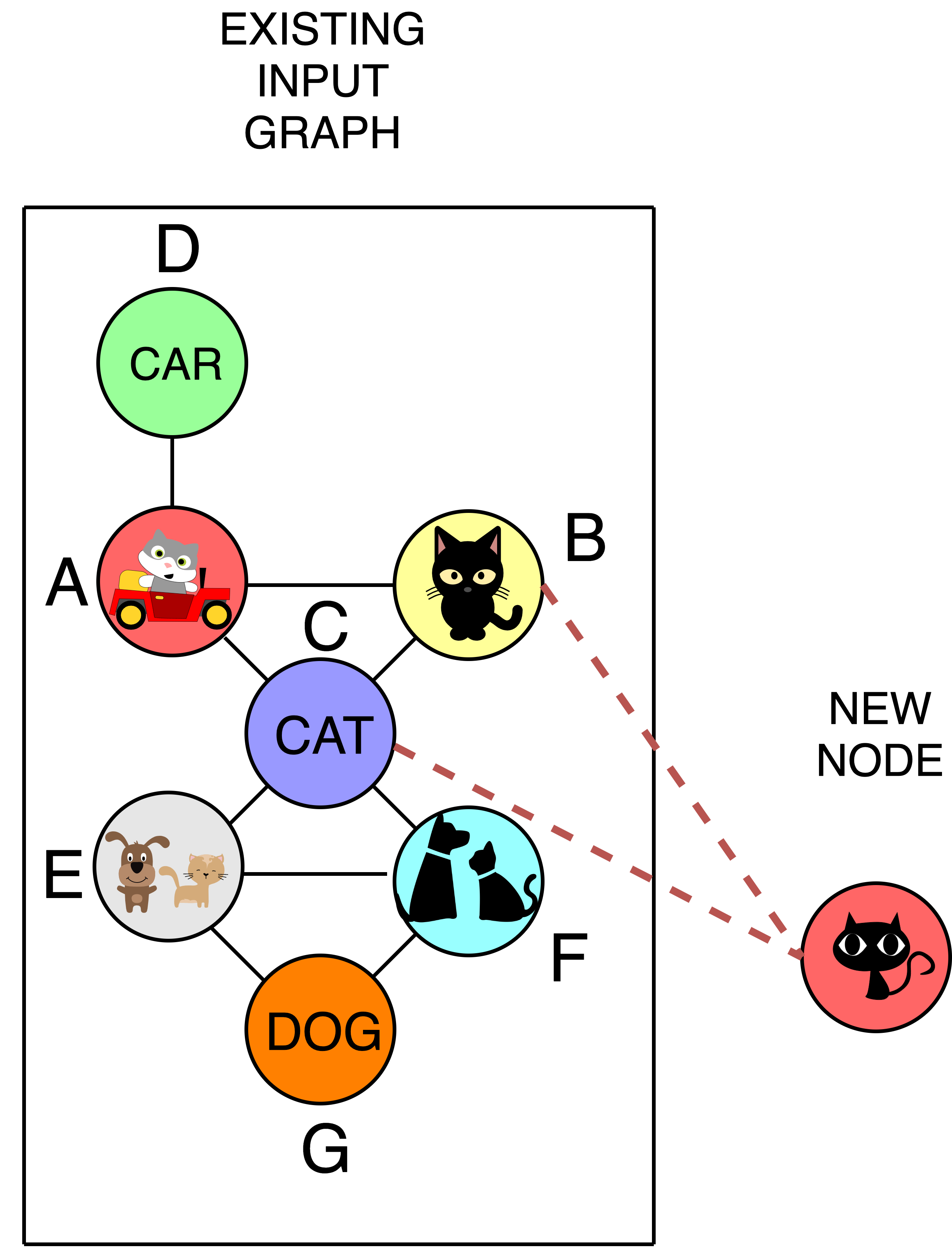}
    \caption{Input-Inference Graph: \normalfont The new node can be connected to the existing graph in various ways, e.g.\  tag-only,  image-only. This provides coarse-grained control over the retrieval results.}
    \label{fig:input-and-inference-graph}
\end{figure}

The overall graph structure that we use is a combination of nodes of different modalities and edges, which establish associations between image tags, image similarity, etc. In our experiments, we connected every image with its corresponding tags and the image’s K nearest neighbors. Image tags  usually refer to objects in the image but can also be any short sequence of words that describe the image contents such as Instagram hashtags or flicker user tags. We get the K nearest neighbors for the query image using the visual similarity score over all the images in the corpus. This ``visual'' similarity is based on an independent deep neural network trained and optimized for representation learning of images in a high-dimensional space where similarity can be captured through a distance metric, i.e.\ closer in distance being more similar. Although we use a ``visual'' similarity metric for neighbor edges in the graph, our framework is generalized to any similarity metric between nodes of the same type or between different node types. In the input graph in \textbf{Figure \ref{fig:unrolled-graph} | Left}, each image node is connected to its visual K-Nearest-Neighbors and its corresponding tag nodes which represent conceptual information present in the image. Image-image and image-tag connections are shown by solid black lines and as shown in \textbf{Figure \ref{fig:input-and-inference-graph}}, new node connections to the existing graph are shown by dotted lines.

\subsection{Training}
After building the above multi-modal graph, we use a training method similar to the one described in GraphSAGE \cite{GraphSAGE}. Our final goal is to convert each node into embeddings (i.e.\ a feature vector) which contains visual information and conceptual information extracted from the graph connections. Once the graph nodes are initialized with any pre-trained embedding,\footnote{The initialization could also be random.}\ we reduce the distance between the node embeddings which are directly connected and increase the distance between the ones that are more than one hop away. 

We take information from the neighborhood of a node and transform it using neural network weights and then aggregate this information. This aggregated information is again transformed and aggregated in a similar fashion and is then used to update the node embeddings for each node. This transformation and aggregation is performed for every node and they share the parameters (weights used to transform).
 \textbf{Figure \ref{fig:unrolled-graph} | Right} shows this process. The black boxes represent the neural networks which help in this aggregation and transformation. This overall method exploits the rich neighborhood information to learn a superior embedding compared to other methods. The training is performed in an unsupervised manner. So, the loss function contains the term for a random walk through which we encode the probability of nodes co-occurring together and hence select negative samples based on those walks. This helps ensure that similar nodes occupy nearby spaces in the embedding space and dissimilar nodes move farther apart. A detailed description of this training can be found in \cite{GraphSAGE}. 

\begin{figure*}
\includegraphics[width=\textwidth]{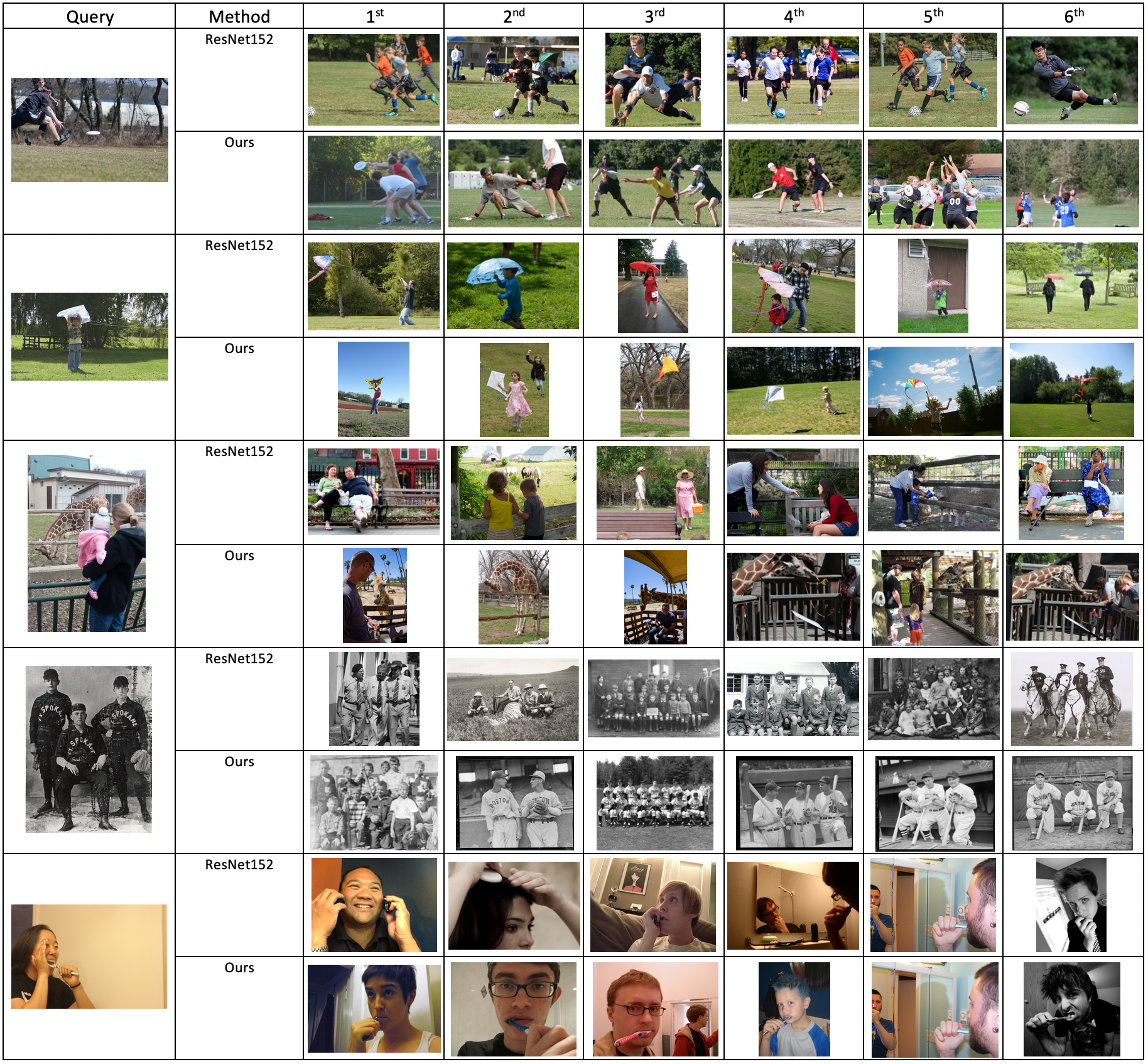}
\caption{ResNet vs.\ Our Method on MS-COCO: \normalfont Our method captures  visual-conceptual nuances which  ResNet cannot. For example, in the 2nd query image, our approach captures the human with a kite and is able to retain the semantic meaning of the scene whereas in ResNet, the kite is lost despite being an important object in the scene. }
\label{resnet-mgnet}
\end{figure*}

\subsection{Inference}
We can perform the following types of inferences with our model, allowing us to power different types of eCommerce search features depending on the user's needs:
%%\begin{figure*}
%%\centering 
%%\includegraphics[width=0.5\textwidth]{images/im2im.png}\includegraphics[width=0.5\textwidth]{images/im2im.png}
%%\caption{a figure with three subfigures}
%%\end{figure*}
\begin{figure*}
    \centering
    \begin{subfigure}[b]{0.19\linewidth}
         \centering
         \includegraphics[width=\linewidth]{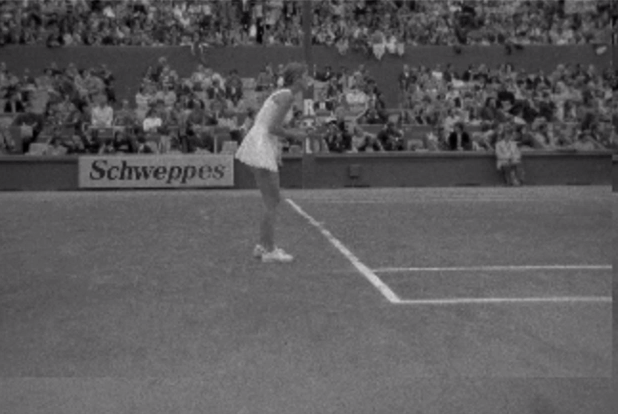}
         \begin{minipage}{.1cm}
            \vspace{3.9em}
            \end{minipage}
         \caption{tennis-query}
         \label{fig:slider-query}
     \end{subfigure}
    \hfill
    \begin{subfigure}[b]{0.19\linewidth}
         \centering
         \includegraphics[width=\linewidth,height=15.2em]{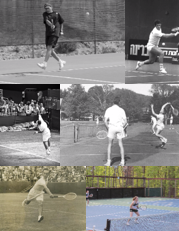}
         \caption{$visualWeight=1.0$}
         \label{fig:weight-one}
     \end{subfigure}
    \hfill
    \begin{subfigure}[b]{0.19\linewidth}
         \centering
         \includegraphics[width=\linewidth,height=15.2em]{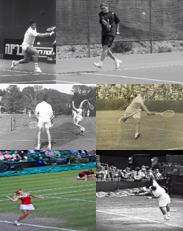}
         \caption{$visualWeight=0.8$}
         \label{fig:weight-zeropointeight}
     \end{subfigure}
    \hfill
    \begin{subfigure}[b]{0.19\linewidth}
         \centering
         \includegraphics[width=\linewidth,height=15.2em]{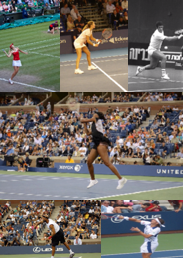}
         \caption{$visualWeight=0.6$}
         \label{fig:weight-zeropointsix}
     \end{subfigure}
    \hfill
    \begin{subfigure}[b]{0.19\linewidth}
         \centering
         \includegraphics[width=\linewidth,height=15.2em]{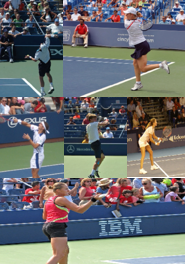}
         \caption{$visualWeight=0.4$}
         \label{fig:weight-zeropointfour}
     \end{subfigure}
    \hfill
        \caption{Visual-Conceptual Control: \normalfont This provides fine-grained control over how much visual-vs.-conceptual similarity  the user wants in the results. Increasing weight  towards the visual features captures the retro effect of the query. Decreasing the reliance on visual features captures a generic tennis scene which perfectly captures the semantics.}
    \label{fig:slider-results}
\end{figure*}

\begin{figure*}
\includegraphics[width=\textwidth]{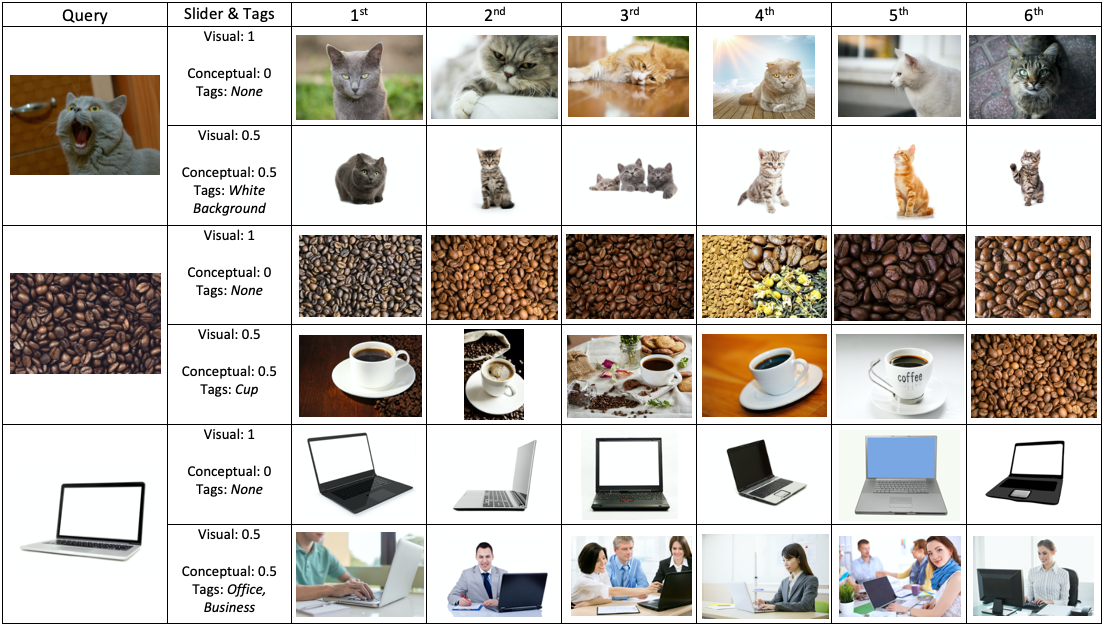}
\caption{Our Method on the Adobe Stock eCommerce site: \normalfont This figure illustrates the several results of our model when trained on Adobe Stock. For example, when  an image-only query is passed as input we get similar images as outputs from our model. Adding a single- or multi-word textual query  with weights captures both the visual and conceptual essence in the results. The degree of inclination towards a particular modality depends on the input weight (see equation \ref{eq:weights}). }
\label{adobe-stock-1}
\end{figure*}

\noindent\textbf{Tag Prediction:} For a given query image, we connect it to the top-k closest images  and then generate a node embedding for it. We can then retrieve the top-k tags based on the least distance with the query node embedding and the tag node embeddings.
\\
\textbf{Image Retrieval:} We have three types of connections that can be used for image retrieval:

\begin{itemize}
    \item \textbf{Only image-image node connections}: When there are no tags available for a query image, we can connect the query image with the top-k closest image nodes using ResNet-152 features or similar embeddings and generate  new image query node embedding. We then retrieve the nearest neighbors using the our model's image node embeddings.
    \item \textbf{Only image-tag node connections:} We can perform the same inference operations with only the query image and its corresponding tag connections.
    \item \textbf{Image-image and image-tag node connections}: For this we connect the query to its nearby image neighbors and also to its corresponding tags and repeat the steps from above. The final graph after connecting the new query image with k=1 is shown in \textbf{Figure \ref{fig:input-and-inference-graph}}.
\end{itemize}

\textbf{Real-Time Slider for Visual-Conceptual Variation in Image Retrieval:} Above we outlined the extreme cases where if we connect the query image only to  image nodes during inference then we will get visually similar images during retrieval, while if we also connect to tag nodes we will get better conceptual image retrieval. To provide more fine-grained control, we use our generated node embeddings for the query image and also its tags. We allow the user to specify the relative weights over the tags and images to reflect their preference for the importance of visual and content similarity. This gives the user fine-grained control over the similarity search so that they can adjust the search results accordingly. To calculate the final embedding we use the following formula:

\begin{figure*}
\includegraphics[width=\textwidth]{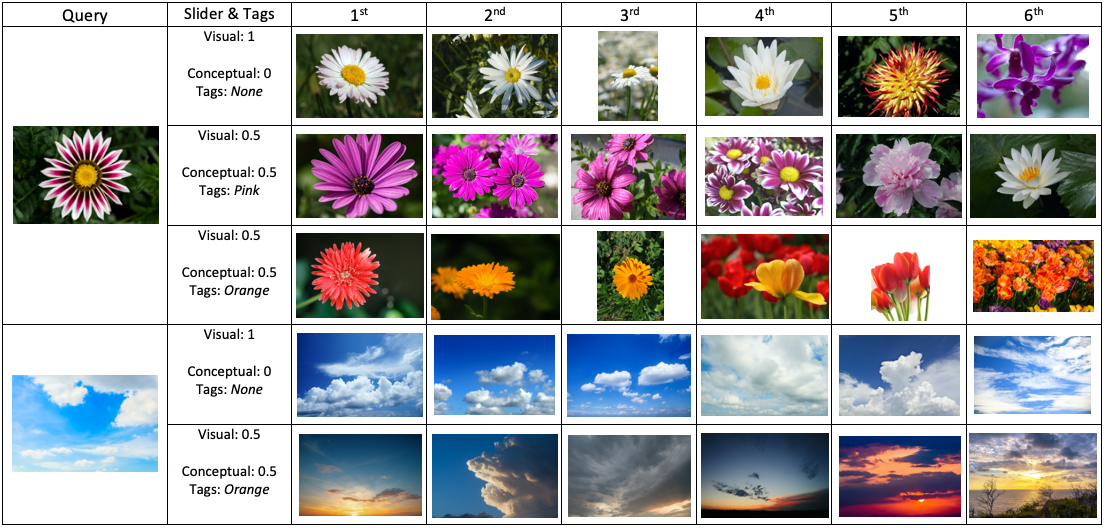}
\caption{Tag Disambiguation: \normalfont This figure illustrates weighted image and tag queries on Adobe Stock. Our Model is able to disambiguate the visual interpretation of a tag with the help of visual cues from the image query. For example, the tag {\em orange} when used in different contexts gives different results.}
\label{tag-disamb}
\end{figure*}

\begin{equation}
\label{eq:weights}
    E = \frac{W_1 \cdot E_i + W_2 \cdot E_t}{2}
\end{equation}
where:

\begin{itemize}
    \item $\;W_1+W_2 = 1$
\item
$E_i =\;$ Our generated node embedding of the query image node
\item
$E_t = \frac{\sum_{i=1}^{total-associated-tags}{E_{t_{i}}}}{number-of-associated-tags} = $ Average of our generated node embeddings of all tag nodes associated with that query 
\end{itemize}

\section{Experiments}
We used MS-COCO for our experiments and our graph had approximately 110K/10K/3K nodes in train/val/test set. Test and val were  restricted to only image nodes. We initialized the image nodes with ResNet-18 features which are of the size 512 and randomly initialize the tag embeddings of  size 512. We selected the top 5 nearest neighbors by using cosine similarity based on ResNet512 features (2048-dimension) and applying a threshold of 0.65.

We trained the model for 50 epochs with a dropout of 20\%, batch size of 512, and a learning rate of 0.00001. We used maximum node degree of 100, negative samples size of 20, model size  ``small'', and model type  ``graphsage\_mean''. First and second layer samples were restricted to 25 and 10 respectively. First and second layer embeddings were 128 dimensions each. Finally, we used a random context to train the model and set the concat parameter to true. This  generated a final embedding of 256 in size. More details regarding these parameters can be found in \cite{GraphSAGE}.

\section{Results}
We first evaluate our framework on MS-COCO quantitatively and then compare it qualitatively with a ResNet152 feature based image retrieval model. We also show qualitative results of how our model performs when weighing visual and conceptual information as would be the case when a user is searching for images on an eCommerce stock image site like Adobe Stock.

\subsection{Quantitative Results}
We used our generated embeddings to extract the top 5 most relevant images for every query image from the unseen test set. A manual, crowd-sourced evaluation  using Figure-8\footnote{www://figure-eight.com}  was performed. The evaluators were asked to rate the retrieved image with respect to the query image along  three criteria:
\begin{enumerate}
        \item Coarse-grained Similarity:
        \begin{enumerate}
            \item Conceptual similarity (terms referring to  background or actions were also counted as concepts).
        \end{enumerate}
        \item Fine-grained Similarity:
        \begin{enumerate}
            \item Objects in the query image appear in retrieved image.
            \item The number of objects matches (e.g.\ single or multiple).
        \end{enumerate}
    \end{enumerate}
\noindent Using these, the evaluators chose one of four relevance labels:
\begin{itemize}
    \item \textbf{Excellent}, if the retrieved image satisfies all  3 criteria based on the query image.
    
    \item \textbf{Good}, if the retrieved image satisfies at least 2 of the 3 similarity criteria. 
    \item \textbf{Acceptable}, if the retrieved image satisfies at least 1 of the 3 similarity criteria
    \item \textbf{Unacceptable}, if the retrieved image does not have any  objects in common with the query image.
\end{itemize}

The results of the evaluation are shown in \textbf{Table \ref{fig:score-dist}}.  Almost 75\% of the images have a score of excellent or good, while less than 10\% are unacceptable. 

\begin{table}[ht]
\centering
\caption{Quantitiative Relevance Results: Score Distribution}
\begin{tabular}{ |c|c|c|c|c| } 
 \hline
 Score & Excellent  & Good & Acceptable & Unacceptable \\
 \hline
 \# of Images & 2164 & 8955 & 2601 & 1275 \\
 \% of Images & 14\% &60\% & 17\% & 9\%\\
 \hline
\end{tabular}
\label{fig:score-dist}
\end{table}

\noindent Using these score we calculate the Normalized Discounted 
Cumulative Gain (nDCG) of the top 5 search results for the image queries.

\begin{comment}
\begin{equation}
    nDCG_p = \frac{DCG_p}{IDCG_p}
\end{equation}
\begin{equation}
    IDCG_p = \sum_{i=1}^{|REL|}{\frac{2^{rel_i}-1}{\log(i+1)}}
\end{equation}

where $|REL|$ is the list of documents ordered by relevance in the corpus up to position $p$.
\end{comment}

The average of nDCG for all queries measures  performance of the ranking algorithm in the search process and reflects key aspects of whole page relevance. Results shown in \textbf{Table \ref{table:ndcg}}.

\begin{table}[ht]
\centering
\caption{Quantitative Relevance Results: nDCG}
\begin{tabular}{|l|l|}
\hline
    nDCG@5: & \textbf{0.39}\\\hline
 nDCG@3: &\textbf{0.41}\\
 \hline
    \end{tabular}
    \label{table:ndcg}
\end{table}

The distribution of the relevance scores shown in \textbf{Table \ref{fig:score-dist}} combined with   the  nDCG scores in \textbf{Table \ref{table:ndcg}} suggest that our model is able to retrieve meaningful images, i.e.\ images that are closely related to the query, but  that it needs further fine-tuning to improve the ranking of the top results.

\begin{comment}
\begin{figure}[h]
\includegraphics[width=0.7\linewidth]{images/figure8distribution.png}
\caption{Score Distribution}
%\label{fig:score-dist}
\end{figure}
\end{comment}

\subsection{Qualitative Evaluation}
Our approach has more control over the conceptual information than  other methods. This means that  during image retrieval we see more  objects that are shared with the query image. We initially trained our model on MS-COCO. \textbf{Figure \ref{resnet-mgnet}} qualitatively compares some results of our method and of using the ResNet152  image retrieval method directly. In general,  ResNet misses some visual cues which  results in changes  to the concepts in the retrieval image compared to the query image. Our model, which the extra information, is able to capture these nuances and maintain the scene concepts while making sure visually relevant images are retrieved. 

As discussed above, due to our unique inference architecture, we are able to retrieve results on different ends of the spectrum (visual-vs.-conceptual) providing fine-grain control  to the  user as to the relative weight given to visual and conceptual features.
For the query image shown in \textbf{Figure\ref{fig:slider-query}}, in a more visually weighted environment, the results  are closer to the query visually showing other retro style tennis images. As we give the concepts of the query image more importance, the results reflect a more generic tennis scene.

We also trained our model on Adobe Stock in order to see how it would perform in an eCommerce environment. The Adobe Stock images have contributor-provided tags, which are generally single words but can be multi-word phrases, and short contributor-provided captions.  Adobe Stock already provides a way for users to search with just keywords, just images, or a combination thereof. However, in the current production search, the keywords always function has hard filters on the tags provided with the images which can result in loss of recall when contributors forgot to provide the relevant tag. \textbf{Figure \ref{adobe-stock-1}} shows results across the Adobe Stock catalog for image search without concepts (conceptual weight=0) and then when a concept is added at weight=0.5. These concepts can be a single keyword, such as {\em cup} in the second row, or multiple keywords, such as {\em white background} in the first row and {\em office business} in the last row. It is particularly interesting that the model can capture abstract concepts such as {\em business} and domain specific concepts such as {\em white background} which is  important in stock photography. In \textbf{Figure \ref{tag-disamb}}, we show how our model is able to distinguish between different  ways that the same embedding is realized in different conceptual spaces by using the information from a different modality. For example, the tag {\em orange} in the textual space represents the color of the flowers when the query image is of a white flower but represents the color of the sky and clouds when the query image is of a blue sky with white clouds. Given the millions of images available on Adobe Stock, providing ways for users to query using combinations of images and keyword terms enables them to smoothly explore this vast inventory to find what they need. 

\section{Conclusions and Future Work}

Most real world applications of image retrieval like Adobe Stock, which is a marketplace for stock photography and illustrations, need a way for users to find visually similar images and to then restrict them based on image concepts. To provide users with fine-grained control over image- and text-based queries, we trained a joint vision and attribute (in this experiment, textual concept tags) embedding in the same high-dimensional space.  We model the visual and attribute relationships as a graph structure, which is able to capture the rich information through node neighborhood. This graph structure helps us  learn multi-modal node embeddings using Graph Neural Networks. We  introduced a novel query-time control, based on selective neighborhood connectivity, allowing the user to control the relative weight of visual and conceptual features to better explore the result set. We evaluates these multi-modal embeddings quantitatively on the downstream relevance task of image retrieval on MS-COCO dataset and qualitatively on MS-COCO and an Adobe Stock dataset.

There are three immediate extensions for improving the quality of the model. First, in order to improve the visual-semantic embeddings, an attention mechanism could be added to focus on certain objects in the image node when the embeddings are updated. Second, we could use  scene-graph embeddings that encode the spatial relationships between objects as our initial point to help our network converge faster. Finally, connecting similar tags could help the network learn better and faster. For example, connecting tags like {\em dog} and {\em puppy} would allow those nodes to share their neighborhood information efficiently.

In this paper we explored multi-modal graphs which jointly embedded image and textual tag features. However, eCommerce catalogs contain additional information such as product category, seller, review ratings, and price. Incorporating this type of information, especially product category, directly into the graph embeddings could provide more accurate search results, whether for stock photo search or more traditional eCommerce search for physical goods.

Finally, there is  future work on how to optimally utilize the multi-modal embeddings in an eCommerce image search application, in our case Adobe Stock. The embeddings trained here could be used directly in the existing search engine. However, this would not fully utilize the underlying capability of the system to let the user smoothly explore the visual-vs.-conceptual importance of their query. To do enable this, the search UX needs to be enhanced to provide user controls beyond the current image similarity with textual filtering.

\begin{acks}
The authors would like to thank Kshitiz Garg for his valuable feedback on our work and Tracy King for reviewing and helping edit this paper.
\end{acks}

%%\begin{figure}
%%\centering 
%%\includegraphics[width=0.45\linewidth,height=8.1em]{images/kite_query.png}\includegraphics[width=0.45\linewidth]{images/frisbee_query.png}
%%\caption{query}
%%\end{figure}

%%\begin{figure}
%%\centering 
%%\includegraphics[width=0.5\linewidth]{images/kite_resnet_results.png}\includegraphics[width=0.5\linewidth,height=13.8em]{images/kite_sensage_results.png}
%%\caption{kite results}
%%\end{figure}

%%\begin{figure}
%%\centering 
%%\includegraphics[width=0.5\linewidth,height=13.8em]{images/frisbee_resnet_results.png}\includegraphics[width=0.5\linewidth,height=13.8em]{images/frisbee_sensage_results.png}
%%\caption{frisbee results}
%%\end{figure}

%%\begin{figure*}
%%\centering 
%%\includegraphics[width=0.5\textwidth]{images/kite_adobe_results_att.png}\includegraphics[width=0.5\textwidth]{images/kite_adobe_results_cont.png}
%%\caption{a figure with three subfigures}
%%\end{figure*}

%%\begin{figure*}
%%\centering 
%%\includegraphics[width=0.5\textwidth]{images/frisbee_adobe_results_att.png}\includegraphics[width=0.5\textwidth]{images/frisbee_adobe_results_cont.png}
%%\caption{a figure with three subfigures}
%%\end{figure*}

% \begin{acks}
% \end{acks}

\bibliographystyle{ACM-Reference-Format}
\bibliography{sample-base}

%%\appendix

%%\section{Research Methods}

%%\subsection{Part One}

\end{document}